\documentclass[preprint, 11pt]{elsarticle}
\usepackage[margin=1.3in]{geometry}
\usepackage{amsmath}
\usepackage{amssymb}
\usepackage{gensymb}
\usepackage{algorithm}
\usepackage[noend]{algpseudocode}
\usepackage{graphicx}
\usepackage{varioref}
\usepackage{color, soul}
\usepackage[labelformat=simple]{subcaption}
\usepackage{multirow}
\usepackage{url}
\usepackage{enumitem}
\usepackage[table,xcdraw]{xcolor}
\usepackage{pgfplots}
\usepackage{setspace}
\usepackage{hyperref}
\usepackage{doi}
\usepackage{xcolor}
\usepackage{float}
\usepackage[normalem]{ulem}

\usepackage{fancyhdr}    % For custom headers/footers

\fancypagestyle{titleheader}{
  \fancyhf{} % clear header and footer
  \fancyhead[C]{%
	\\[-8ex]%
	{\color{red}\small\textbf{This is a pre-print of an article published in the European Journal of Operational Research: \url{https://doi.org/10.1016/j.ejor.2021.11.031}}}\\[1ex]%
    {\color{red}\small\textbf{Update:}}~%
    {\color{red}\small We have released an open-source version of the algorithm at \url{https://github.com/JeroenGar/gdrr-2bp}}
  }
}

\usepackage[draft]{todonotes}   % notes showed

\DeclareMathSymbol{\shortminus}{\mathbin}{AMSa}{"39}

\makeatletter
\def\BState{\State\hskip-\ALG@thistlm}
\makeatother

\journal{European Journal of Operational Research}

\begin{document}
\begin{frontmatter}
	\title{A Goal-Driven Ruin and Recreate Heuristic for the 2D Variable-Sized Bin Packing Problem with Guillotine Constraints}
	
	\author[1]{Jeroen Gardeyn\corref{cor1}}
	\ead{jeroen.gardeyn@kuleuven.be}
	
	\author[1]{Tony Wauters}
	\ead{tony.wauters@kuleuven.be}
	
	\cortext[cor1]{Corresponding author}
	
	\address[1]{KU Leuven, Department of Computer Science, NUMA, Belgium}

	\begin{abstract}
		This paper addresses the two-dimensional bin packing problem with guillotine constraints. 
		The problem requires a set of rectangular items to be cut from larger rectangles, known as bins, while only making use of edge-to-edge (guillotine) cuts. 
		The goal is to minimize the total bin area needed to cut all required items. 
		This paper also addresses variants of the problem which permit 90$\degree$ rotation of items and/or a heterogeneous set of bins.
		A novel heuristic is introduced which is based on the ruin and recreate paradigm combined with a goal-driven approach.
		When applying the proposed heuristic to benchmark instances from the literature, it outperforms the current state-of-the-art algorithms in terms of solution quality for all variants of the problem considered.
		
	\end{abstract}
\end{frontmatter}

\section{Introduction}
\thispagestyle{titleheader}
The Two-Dimensional Bin Packing Problem (2BP) consists of packing a  heterogeneous set of small rectangular items into larger rectangular bins.
In general, a 2BP solution is considered feasible if (1) the items are fully inside the bin, (2) items are not overlapping and (3) the edges of the items are parallel to the edges of the bin. The goal is to minimize the amount of unused bin area incurred by packing all the items. 

\citet{lodi1999heuristic} defined four variants of the 2BP problem using a three-field notation. This notation defines whether the items are allowed to rotate 90 degrees and/or whether guillotine cuts are required.
This paper focuses on variants of the 2BP which require guillotine cuts (2BP\textbar *\textbar G). Both fixed orientation (2BP\textbar O\textbar G) and the variant that allows 90$\degree$ rotation of items (2BP\textbar R\textbar G) are addressed.

\begin{figure}[htb!]
	
	\centering
	\begin{subfigure}{.4\textwidth}
		\centering
		\includegraphics[width=0.5\linewidth]{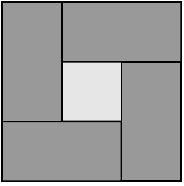}
		\caption{Non-guillotine pattern}
		\label{fig:guillotine1}
	\end{subfigure}%
	\begin{subfigure}{.4\textwidth}
		\centering
		\includegraphics[width=0.5\linewidth]{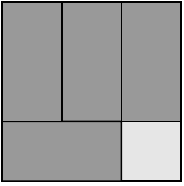}
		\caption{Guillotine pattern}
		\label{fig:guillotine2}
	\end{subfigure}
	\caption{Visual representation of non-guillotine and guillotine patterns}
	\label{fig:guillotine}
\end{figure}

The guillotine constraint defines that all cuts must be edge-to-edge cuts.
Figure \ref{fig:guillotine1} shows a simple visual example of a pattern which does not comply with the guillotine constraint. It is not possible to cut this pattern by exclusively using edge-to-edge cuts. The pattern in Figure \ref{fig:guillotine2}, on the other hand, satisfies the guillotine constraint by first making a horizontal cut followed by several vertical cuts.
Guillotine cuts have many real-world applications as they are often required in glass- and wood-cutting industries.

This work introduces a novel heuristic based on the ruin and recreate paradigm combined with a goal-driven approach (GDRR). 
The heuristic attempts to improve the solution at each iteration by removing and reinserting items into the bins in a greedy fashion.
GDRR is goal-driven in the sense that it iteratively lowers the available bin area. 
Each time the algorithm finds a feasible solution with a certain total bin area, it is henceforth forced to find new solutions which use less bin area. 
As a result, most of the time, the heuristic will be unable to fit all items into the available bins and will therefore be working with infeasible solutions. 
After each run, items which could not be placed are automatically reconsidered for reinsertion during the next run. In order to reach complete solutions, the objective function penalizes unassigned items in order to steer the heuristic towards feasibility. 

GDRR is also capable of addressing the variant with a heterogeneous set of bins. This generalization is commonly referred to as the 2BP with variable-sized bins. Since the classification made by \citet{lodi1999heuristic} does not cover this variant, it will henceforth be referred to as the 2VSBP.

In summary, the proposed heuristic addresses four variants of the 2BP:
\begin{itemize}
	\item \textbf{2BP\textbar O\textbar G}: identical bins, no rotation
	\item \textbf{2BP\textbar R\textbar G}: identical bins, 90$\degree$ rotation
	\item \textbf{2VSBP\textbar O\textbar G}: heterogeneous set of bins (variable-sized), no rotation
	\item \textbf{2VSBP\textbar R\textbar G}: heterogeneous set of bins (variable-sized),  90$\degree$ rotation
\end{itemize}
The remainder of the paper is organized as follows.
In Section \ref{section:literaturereview} a compact literature review is provided which includes papers covering both the traditional 2BP and the variant with a heterogeneous set of bins. Section \ref{section:datastructure} introduces the data structure for solution representation. Next, a detailed explanation of the complete ruin and recreate heuristic is given in Section \ref{section:ruinandrecreateheuristic}. A comparison of the heuristic against the current state of the art is conducted in Section \ref{section:computationalresults}. Finally, Section \ref{section:conclusion} summarizes the paper and highlights some future research directions.

\section{Literature review} \label{section:literaturereview}
General typologies for cutting and packing problems are published in \citet{Dyckhoff1990} and \citet{Wascher2007}. Surveys for the 2BP have been provided by \citet{Lodi2002} and \citet{lodi2002two}. These surveys cover heuristics, exact methods and bounds for the 2BP.
A more recent survey focusing on exact methods was conducted by \citet{iori2020exact}. 

\citet{gilmore1965multistage} were the first to tackle the 2BP and proposed a column generation approach for the multidimensional bin packing problem.
Many heuristics for 2BP\textbar *\textbar G have since been published in the literature and this review will focus on the most competitive among these.
\citet{Polyakovsky2009} developed a 'guillotine bottom left' constructive heuristic to pack one bin at a time in combination with a pseudo-parallel agent-based system wherein each agent has its own characteristics.
Later, \citet{Charalambous2011} introduced a constructive heuristic which supports item rotation. At each iteration, their method packs one bin by creating multiple simple cutting patterns and selects the one with the highest quality by way of a sufficiency criterion. To improve solution quality, they proposed a post-optimization routine which prioritizes items that most often do not fit during the constructive phase.
Later, \citet{fleszar2013three} proposed three new insertion heuristics and a justification improvement heuristic.
\citet{Lodi2017} introduced a heuristic based on an enumeration tree for the 2BP\textbar O\textbar G. At each level of the tree, a new bin is packed. When filling a new bin, a set of selection and guillotine split rules are considered. These different strategies create new nodes in the tree. In an enhanced version of the heuristic, they attempt to reduce the search space to improve performance.
More recently, \citet{Cui2015} present a sequential value correction heuristic which repeatedly creates cutting patterns while trying to maximize the total item value. The values of the items are updated after each iteration based on information from current and past solutions.
Finally, a heuristic which uses a pattern generation procedure that creates triple-block patterns was introduced by \citet{Cui2018}. The procedure is followed by an improvement phase which uses an Integer Linear Programming model to select the best patterns.

The concept of heterogeneous sets of bins (variable-sized) was first introduced by \citet{friesen1986variable}. They proposed three heuristics for one-dimensional bin packing problems.
Literature on the 2VSBP\textbar *\textbar G is more scarce compared to the variant with identical bins. To the best of our knowledge, only the variant without rotation of items has been studied. \citet{ortmann2010new} proposed a two-stage heuristic, which begins with a strip packing procedure. Next, attempts are made to repack these strips into smaller and smaller bins.
Later, \citet{Hong2014} designed a hybrid heuristic for the 2VSBP\textbar O\textbar G. They proposed a mixed bin packing algorithm which is used in combination with iterative simulated annealing runs and a binary search.
Finally, \citet{wei2013goal} introduced a heuristic with a goal-driven approach for the 2VSBP\textbar O\textbar F (no guillotine constraint). Their algorithm packs bins with a sequential packing heuristic, followed by a local search to improve the solution. It also performs a binary search on bin combinations and increases the search effort each time.

It should be clear that many of these solution methods are either based on exact methods or focus heavily on the constructive aspect. Exact methods, however, often lack the flexibility required to adapt to real-world scenarios and generally do not scale well for larger problems. 
On the other hand, although constructive heuristics are usually more flexible and very fast, their solutions are often lacking in terms of quality.
The approach proposed in this paper attempts to find a balance by focusing on heuristic improvement strategies.

\section{Solution representation}\label{section:datastructure}
\begin{figure}[h]
	\centering
	\includegraphics[width=0.8\textwidth]{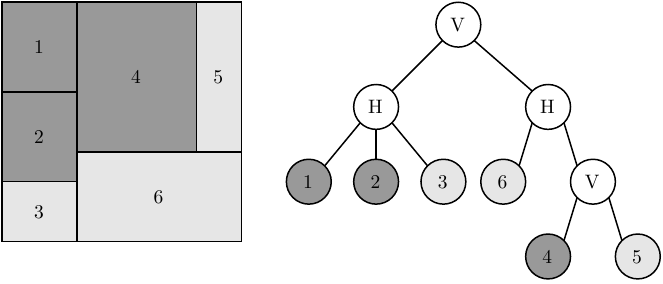}
	\caption{Visual representation of the data structure}
	\label{fig:datastructure}
\end{figure}

Each cutting pattern is represented as a rooted tree, as shown in Figure \ref{fig:datastructure}. The tree consists of three different types of node. 
There are item (dark gray), leftover (light gray) and structure nodes (white).
Leaves are always either an item or leftover node. All non-leaf nodes define the structure of a cutting pattern in a hierarchical way and are denoted as either vertical (V) or horizontal (H). The orientation of the node corresponds to the direction in which its children are cut. This orientation is identical for every node of a level and each new level switches the orientation compared to the previous one.

A similar, but not identical, data structure was used before by \citet{clautiaux2013new}, \citet{fleszar2013three} and \citet{kroger1995guillotineable}.
As noted by \citet{fleszar2013three}, this representation always satisfies the guillotine constraint and never allows for overlapping items.

Note that the data structure does not define any coordinates. It only describes relative positions of items and leftovers.
This allows for simpler insertion or removal of items (or groups of items) compared to more rigid structures where a cutting pattern is defined by a set of items and their locations.

As was first described by \citet{gilmore1965multistage}, in real-world scenarios there is often a limit on the number of possible stages. 
A stage includes one or more horizontal or vertical cuts, but never a mixture of the two. 
Progressing from one stage to another involves rotating the cut orientation. 
The number of stages therefore represents the number of blade rotations needed to cut the pattern. 
In this representation, since the cutting orientations are switched at every level, each level represents a stage.
While not the focus of this paper, limiting the number of stages in this data structure involves simply defining a tree's maximum depth.

Due to the nature of the heuristic, there is a need to be able to represent incomplete solutions. Therefore, a solution $S=\{C, E\}$ not only contains a set of cutting patterns $C$, it also consists of a set of excluded items $E$. A solution is deemed feasible if all items are included ($E=\varnothing$).

\section{Ruin and recreate heuristic} \label{section:ruinandrecreateheuristic}
\begin{figure}[!htb]
	\centering
	\includegraphics[width=0.8\textwidth]{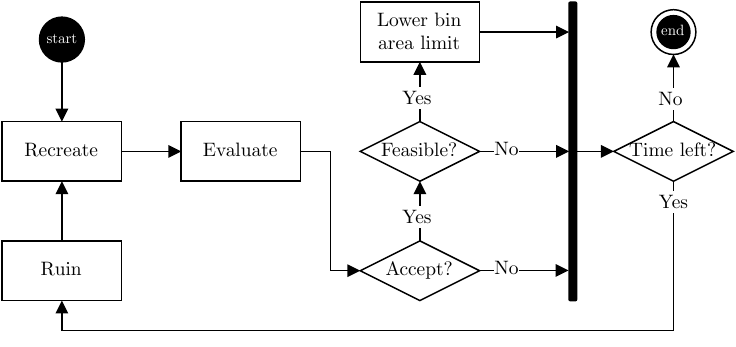}
	\caption{High-level overview of GDRR}
	\label{fig:generaloverview}
\end{figure}

Figure \ref{fig:generaloverview} provides a high-level overview of the heuristic. 
At each iteration of the local search, the solution is partially destroyed and rebuilt in an attempt to find improvements.
Every time a new feasible solution is found, the bin area limit is reduced. The general idea behind this is discussed in Section \ref{section:GDA}. 	
Detailed explanations of the ruin and recreate procedures are given in Sections \ref{section:ruin} and \ref{section:recreate}. Afterwards, the modified solution is evaluated and has the possibility of being accepted. This is covered in Section \ref{section:evaluation}.
The entire process is repeated until a given time limit is reached.

\subsection{Goal-driven approach} \label{section:GDA}
In the 2BP, the quality of a solution is based on the sum of areas of the bins required to pack all items. Due to the nature of the problem, all feasible solutions which result in the same total bin area are deemed to be equal in quality. This means that, once a complete solution is found with a certain total bin area, solutions with equal (or greater) total bin area are no longer of interest.

\citet{wei2013goal} proposed a goal-driven approach (GDA) heuristic for the 2BP with variable-sized bins. It generates all combinations of bins that have a total area between a lower and upper bound, sorts them by area and attempts to find the best combination that produces a feasible solution. The sorted list is traversed using a binary search. At each iteration, the GDA heuristic must use that specific combination of bins. Although this particular algorithm does not incorporate the guillotine constraint, the general idea is applicable.

Similar to GDA, the proposed approach could also be viewed as being goal-driven. 
The recreate phase, responsible for rebuilding the ruined solution, must adhere to a certain bin area limit.
A new bin can only be opened if the total bin area of the solution remains below this threshold.
Therefore, most of the time, the recreate phase will be unable to fit all items into the available bins and will produce infeasible solutions.
A cost function guides the heuristic towards a solution capable of fitting all items.

\begin{figure*}[htb]
	\centering
	\includegraphics[width=0.8\textwidth]{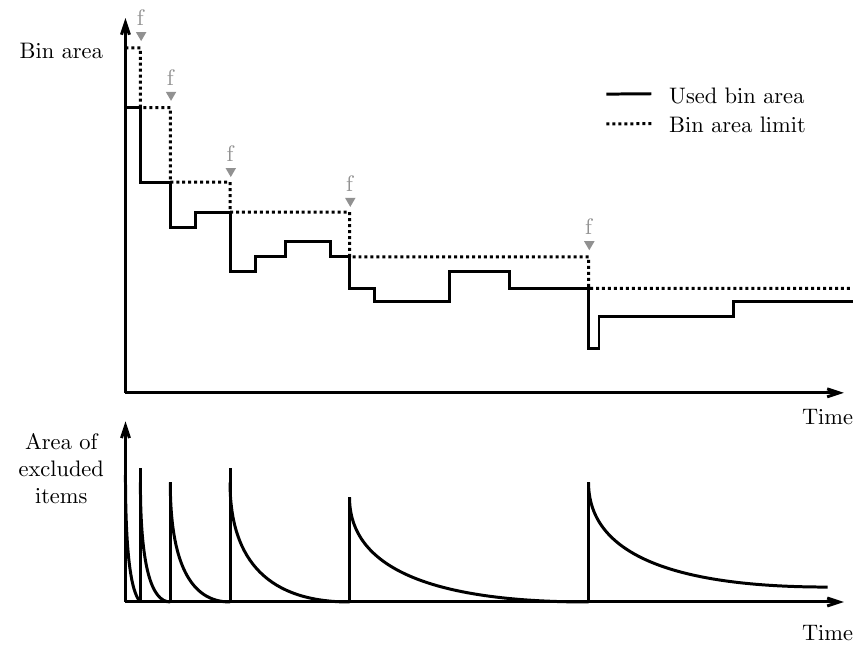}
	\caption{Progression of GDRR over time}
	\label{fig:gda}
\end{figure*}

Figure \ref{fig:gda} represents one possible course of the heuristic over time.
At all times, the goal is to find a set of cutting patterns capable of packing all items with a total bin area below the current limit. 
Whenever a new feasible solution (denoted by `f') is found, the limit is lowered to the total bin area of this new solution.
A formal description of the complete GDRR heuristic and how the bin area limit tightens is presented in Algorithm \ref{alg:GDRR} at the end of Section \ref{section:ruinandrecreateheuristic}.	

Unlike \citet{wei2013goal}, the search does not begin from scratch each time, but rather the solution is destroyed until the bin area constraint is once again satisfied.
Therefore, high-quality features from a previous solution can be carried over when the limit is lowered. This can result in a significant amount of saved computational time. However, the heuristic may be more likely to get stuck in local optima when solutions are already tightly packed.

Another difference lies in the combination of available bins. GDA tries to make a feasible solution with a fixed combination of bins. By contrast, GDRR is free to use whichever combination of bins yields good results according to the cost function. GDRR only needs to ensure that the total bin area is below the current limit.
Furthermore, it can sometimes happen that there is a feasible solution for a combination of bins $A$, but not for a combination of bins $B$ despite the total bin area of $A$ being smaller than $B$.
This situation most often occurs with instances that contain small bins (few items per bin) or items with extreme dimensions.
Due to the binary search nature of GDA, these combinations might not be reachable.

It should become clear that, by design, the heuristic will be working with infeasible solutions most of the time.
This might seem counterproductive at first, but such a strategy has two major advantages compared to incorporating bin area into a cost function.
First of all, items which are not included in the solution have a new chance at being inserted every iteration. 
This drastically increases the likelihood of finding better solutions. 
If this were not the case, reaching complete solutions with less bin area would be unlikely, especially when the cutting patterns are complex.
Second, as shown in Figure \ref{fig:gda}, the total area of excluded items can be used as a very gradual indicator of a solution's quality. This is in stark contrast to the abrupt drops when using bin area as a cost factor.

In conclusion, although the proposed heuristic is also goal-driven, it significantly differs from the approach introduced by \citet{wei2013goal}. They merely share the same philosophy of constraining the use of bins in some way. 		

\subsection{Ruin} \label{section:ruin}

The ruin procedure removes a number of nodes from the solution. All item and structure nodes are candidates for removal. Leftover nodes are ineligible since removing them would not have any effect. They can only be deleted as a result of removing a parent node.

\begin{figure*}[h]
	\centering
	\includegraphics[width=0.8\textwidth]{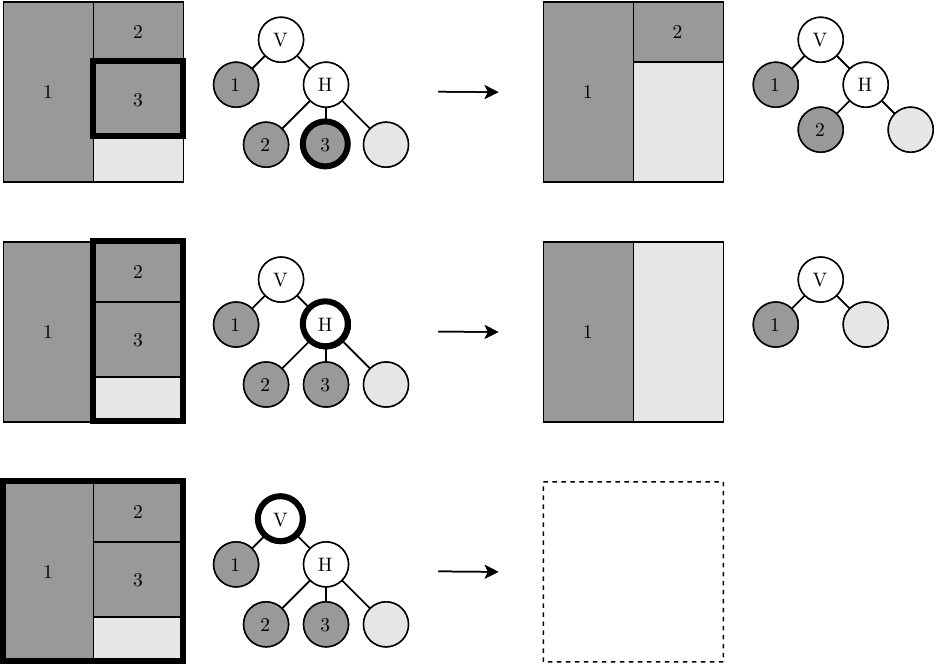}
	\caption{Three examples of node removal}
	\label{fig:ruin}
\end{figure*}

Three cases of node removal are shown in Figure \ref{fig:ruin}. 
These figures illustrate the situation before (left) and after (right) the removal of a node. In each example, the node with a bold edge is the one selected for removal.
In the first example, the node containing item 3 is removed. As a result, the leftover node is expanded to account for the freed-up space. In the second example, a first-level structure node is deleted. This results in items 2 and 3 being removed since they are children. The final example shows the root node being selected for removal. This results in the entire cutting pattern being removed. 
It should be clear that impact of a single removal can range from individual items to entire cutting patterns.
The closer a structure node is to the root of the tree, the more significant an impact its removal will have.

There are two main motivations for allowing structure nodes to be removed.
First, removing these nodes produces large regions of contiguous leftover space.
This, in turn, provides the recreation phase more flexibility, resulting in diverse neighboring solutions. 
The second main advantage lies in the fact that it is probable that some of the removed items will form good patterns again at another location. In the second example, for instance, items 2 and 3 are likely to form a compact pattern again because they share the same width.
This is particularly apparent in the later stages of the search, when the solution is already quite good in terms of quality. It could be viewed as a form of related removal. 

However, dramatically destroying the solution can hinder the chances of finding better solutions. Radically ruining a solution too often can result in a lot of wasted time.
Because the most impactful nodes are closest to the root of the tree, they are less numerous compared to the less impactful nodes at the bottom of the tree. Due to the fact that the ruin method selects nodes at random, removals with a significant impact have a lower chance of occurring. This could be viewed as a sort of built-in balancing mechanism.

\begin{algorithm}[htb!]
			\caption{Ruin phase}\label{alg:ruin}
			\begin{tabbing}
				\hspace*{\algorithmicindent} \textbf{Input:} \hspace*{2mm} \= A solution $S = \{C, E\}$ where $C$ is a set of cutting patterns\\ 
				\> and $E$ is a set of excluded items, bin area limit $\mathcal A_{lim}$, and \\
				\> the average number of nodes to be removed $\mu$ \\
				\hspace*{\algorithmicindent} \textbf{Output:} \> A modified (ruined) solution $S'$
			\end{tabbing}
			\begin{algorithmic}[1]
				\Function{ruin}{$S, \mathcal A_{lim}, \mu$}
				\State $S' = \{C', E'\} \gets S$
				\State $i \in_R \{1,...,2\mu-1\}$ 
				\While{$i > 0 \textbf{ or } \sum_{c \in C'} \mathcal A_{c} \geq \mathcal A_{lim} $}
				\State $c \in_R C'$
				\State $N_c \gets \text{item and structure nodes of } c $
				\State $n \in_R N_c$
				\State $c' \gets \textsc{remove}(c,n)$
				\State $C' \gets C' \setminus \{c\}$
				\If{$c' \text{ is not empty}$}
					\State $C' \gets C' \cup \{c'\}$
				\EndIf
				\State $E_n \gets \text{items in } n$
				\State $E' \gets E' \cup E_{n}$
				\State $i \gets i-1$
				\EndWhile
				\State \Return $S'$
				\EndFunction
			\end{algorithmic}
		\end{algorithm}

Algorithm \ref{alg:ruin} provides the pseudocode for the ruin phase. 
The number of nodes to be removed, $i$, is uniformly randomized ($\in_R$) at each iteration and averages out at $\mu$ (line 3). 
The procedure continues ruining until $i$ nodes have been removed or while the total bin area of the modified solution exceeds the bin area limit $\mathcal{A}_{lim}$ (line 4). 
For each removal, a cutting pattern $c$ is selected uniformly at random (line 5). 
A random item or structure node $n$ is selected from $c$ (lines 6-7). 
This node is removed from $c$, resulting in $c'$ (lines 8-9). 
Finally, both the set of cutting patterns and the set of excluded items associated with the modified solution $S'$ are updated (lines 10-13).

\subsection{Recreate} \label{section:recreate}

During the recreate phase, the (ruined) solution needs to be rebuilt. This is done by inserting items into either leftover nodes or new bins.
Of course, due to the bin area limit, it is highly unlikely that it will be possible to insert all items. 

\begin{algorithm}[htb!]
	\caption{Recreate phase}\label{alg:recreate}
	\begin{tabbing}
		\hspace*{\algorithmicindent} \textbf{Input:} \hspace*{2mm} \= A solution $S = \{C, E\}$ where $C$ is a set of cutting patterns\\ 
		\> and $E$ is a set of excluded items, bin area limit $\mathcal A_{lim}$ \\
		\hspace*{\algorithmicindent} \textbf{Output:} \> A modified (recreated) solution $S'$
	\end{tabbing}
	\begin{algorithmic}[1]
		\Function{recreate}{$S, \mathcal A_{lim}$}
		\State $S' = \{C', E'\} \gets S$
		\State $\tilde{E} \gets E'$
		\While{$\tilde{E} \ne $ \O}
		\State $e_{mr} \gets \text{most restricted item} \in \tilde{E}$
		\State $O \gets \text{insertion options for } e_{mr} \text{ for all cutting patterns}\in C'$
		\If {$O = $ \O} $\color{cyan} \texttt{ //attempt to open new bin}$
			\State $\Delta \gets \mathcal A_{lim} - \sum_{c \in C'} \mathcal A_c$ 
			\If{$\text{there is a bin with } \mathcal A < \Delta$}
				\State $c \gets \text{empty cutting pattern using random bin with } \mathcal A_c < \Delta$
				\State $C' \gets C' \cup \{c\}$
				\State $O \gets O \cup \{\text{insertion options for } e_{mr} \text{ in } c\}$
			\EndIf
		\EndIf
		\If {$O \ne $ \O} $\color{cyan} \texttt{ //insert the item}$
			\State $o \gets \text{select best (with blinks)} \in O$
			\State $c \gets \text{cutting pattern of } o$
			\State $c' \gets \textsc{insert}(c,o)$
			\State $C' \gets C' \setminus \{c\} \cup \{c'\}$
			\State $E' \gets E' \setminus \{e_{mr}\}$
		\EndIf
		\State $\tilde{E} \gets \tilde{E} \setminus \{e_{mr}\}$
		\EndWhile
		\State \Return $S'$
		\EndFunction
	\end{algorithmic}
\end{algorithm}

Algorithm \ref{alg:recreate} provides the pseudocode for the recreate phase.
During this phase, all currently excluded items have the opportunity to be inserted into the solution (lines 3-4). Set $\tilde{E}$ corresponds to all items which have yet to receive a chance at insertion. 
At each iteration, the most restricted item is selected as the next insertion candidate (line 5).
The restrictiveness of an item is determined by its number of possible insertion places: the more possibilities, the less restricted\footnote{Determining the most restricted item can be performed in linear time ($O(n)$) by caching the insertion options for all excluded items in a table and performing updates throughout the recreate phase.}. When multiple items are deemed equally restricted, a random item is selected. The definition of insertion options is provided in Section \ref{section:insertionoptions} in addition to how their costs are calculated.
For the selected item, $e_{mr}$, all insertion options are generated (line 6).
If there are no available options (line 7), an attempt is made to open a new bin (lines 8-12).
Only bins which will not result in exceeding the bin area limit $\mathcal{A}_{lim}$ are eligible (lines 8-9).
If possible, a new and empty cutting pattern is created with a random eligible bin (lines 10-11) and the set of insertion options is updated (line 12).
If the item can be inserted (line 13), all options are evaluated (line 14) and eventually the item is inserted into the solution (lines 15-18). Most of the time, the cheapest option is selected in a greedy fashion, but sometimes the best option are blinked over.
The concept of blinks is explained in Section \ref{section:blinks}.
Regardless of whether a successful insertion takes place, item $e_{mr}$ is removed from $\tilde{E}$ as it has now received a chance at insertion (line 19). 
Once every item has been given a chance to be inserted, a recreated solution $S'$ is returned (line 20).

\subsubsection{Insertion Options} \label{section:insertionoptions}
\begin{figure}[htb]
	\centering
	\begin{subfigure}{.4\textwidth}
		\centering
		\includegraphics[width=0.8\linewidth]{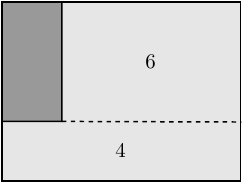}
		\caption{$cost = 12^2 - (6^2 + 4^2) = 92$}
		\label{fig:proposals1}
	\end{subfigure}%
	\begin{subfigure}{.4\textwidth}
		\centering
		\includegraphics[width=0.8\linewidth]{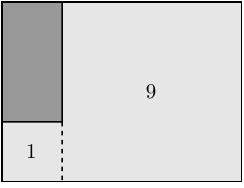}
		\caption{$cost = 12^2 - (9^2 + 1^2) = 62$}
		\label{fig:proposals2}
	\end{subfigure}
	\\
	\begin{subfigure}{.4\textwidth}
		\centering
		\includegraphics[width=0.8\linewidth]{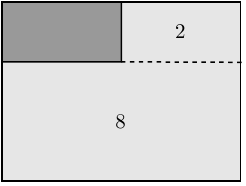}
		\caption{$cost = 12^2 - (8^2 + 2^2) = 74$}
		\label{fig:proposals3}
	\end{subfigure}
	\begin{subfigure}{.4\textwidth}
		\centering
		\includegraphics[width=0.8\linewidth]{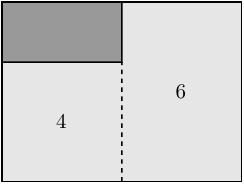}
		\caption{$cost = 12^2 - (6^2 + 4^2) = 92$}
		\label{fig:proposals4}
	\end{subfigure}
	\caption{Four options of inserting an item into a leftover node ($width_{n^u_o} = 4$, $height_{n^u_o} = 3$, $\alpha = 2$, 2BP\textbar R\textbar G).}
	
	\label{fig:proposals}
\end{figure}

An insertion option defines how exactly an item is inserted into a leftover node.
If 90$\degree$ rotation is allowed, there are two possible orientations for each item.
Due to the guillotine constraint, each feasible orientation of an item in a leftover node, in turn, generates two insertion options depending on the direction of the first cut (vertical or horizontal).
As a result, there are at most four different ways of inserting an item into a leftover node. 
An illustrative example of these four possibilities is shown in Figure \ref{fig:proposals}.
For each excluded item, all leftover nodes capable of fitting the item are considered.

All insertion options have an associated cost.
This cost is used as a measure of an option's desirability and is calculated using Equation \ref{eq:costinsertion}.
Inserting items not only consumes existing leftover nodes, but also generates new ones.
The cost of an insertion option $o$ corresponds to the difference in total leftover value before and after the insertion.
Each leftover node thus has a corresponding value (Equation \ref{eq:leftovervalue}) and the aim is to minimize the loss in value resulting from insertion.

\begin{equation*}
    \begin{array}{ll}
        n^{u}_{o} & \text{: leftover node used by insertion option } o           \\
        N^{c}_{o}   & \text{: set of leftover nodes created by insertion option }o \\
        \alpha      & \text{: leftover area value exponent}                        \\
    \end{array} \\
\end{equation*}
\begin{equation}\label{eq:costinsertion}
    cost(o) = value(n^{u}_{o}) - \sum\limits_{n \in N^{c}_{o}} value(n)
\end{equation}
\begin{equation}\label{eq:leftovervalue}
    value(n) = (width_n \cdot height_n)^\alpha
\end{equation}

As shown in Equation \ref{eq:leftovervalue}, the value of a leftover node increases polynomially with respect to area.
This is to encourage the heuristic to preserve large contiguous leftover areas instead of many smaller ones.
Generally speaking, it will be easier to pack items into few, but larger, leftover nodes rather than many small ones.
In Figure \ref{fig:proposals}, option (b) has the lowest cost since it preserves the largest leftover value.

The value of exponent $\alpha$ has an impact on the behavior of the algorithm because it defines the ordering of insertion options. Making this value too large can lead to the creation of very few well-sized leftover areas. It will ignore many medium-sized leftover nodes and only care about a few large ones.
On the other hand, setting the value of $\alpha$ too small can be detrimental for preserving large leftover areas. 

\subsubsection{Blinks} \label{section:blinks}
Always selecting the best option would result in a deterministic recreate phase and may lead to making the same bad decisions over and over again.
To combat this, it is sometimes necessary to turn a blind eye to an insertion option. 
The concept of blinks was first introduced by \citet{Christiaens2020} in a state-of-the-art heuristic for solving Vehicle Routing Problems. 
The general idea, however, is also applicable here. 

In a pure best-fit, the best option is always selected. But with blinks, there is a small chance for an option to be ignored. Blinking rate $\beta$ defines the likeliness of skipping over an option. Equation \ref{eq:blink} shows the chance of an option being selected based on its rank $r$. A value for $\beta$ of $0.05$, for example, means each option has a 5\% chance of being ignored.

\begin{equation} \label{eq:blink}
	p(r) = (1 - \beta)\beta^{(r-1)} \quad\quad r \in \{1,...,\infty\}
\end{equation}
Blinking exhibits the same behavior as Heuristic-Biased Stochastic Sampling, introduced by \citet{Bresina1996}, with an exponential bias function. However, as \citet{Christiaens2020} note, blinking is more efficient since there is no need to sort options based on their fitness.

\subsection{Evaluation} \label{section:evaluation}
After each ruin and recreate iteration, the solution needs to be evaluated. 
This is accomplished by determining the quality of the solution in question, as explained in Section \ref{section:solutionquality}. Based on this quality, the solution is either accepted or rejected using the Late-Acceptance Hill-Climbing metaheuristic, which is described in Section \ref{section:lahc}.

\subsubsection{Solution Quality} \label{section:solutionquality}
Due to bin area limit constraints, the recreate phase will most likely be unable to incorporate all items into the solution.
The goal, therefore, is to guide the heuristic towards a feasible solution.
Equation \ref{eq:quality} describes how two solutions are compared against each other.

\begin{equation} \label{eq:quality}
    \begin{array}{l}
        \begin{array}{ll}
            E_S    & \text{: set of excluded items in solution }S                            \\
            N^{l}_S    & \text{: set of leftover nodes from cutting patterns in $C$ in solution }S \\
            a_e(S) & \text{:$\sum\limits_{e \in E_S} width_e \cdot height_e$}                \\
            v_l(S) & \text{:$\sum\limits_{n \in N^{l}_S} value(n)$}                            \\
        \end{array} \\
        compare(S_1, S_2) = \left\{
        \begin{array}{cll}
            \shortminus1        & \quad \text{if } a_e(S_1) < a_e(S_2)        & $\color{cyan} \texttt{ //S1 superior}$                   \\
            1                   & \quad \text{if } a_e(S_1) > a_e(S_2)        & $\color{cyan} \texttt{ //S2 superior}$                   \\
            \shortminus1        & \quad \text{if } a_e(S_1) = a_e(S_2) \text{ and } v_l(S_1) > v_l(S_2)    & $\color{cyan} \texttt{ //S1 superior}$ \\
            1                   & \quad \text{if } a_e(S_1) = a_e(S_2) \text{ and } v_l(S_1) < v_l(S_2)    & $\color{cyan} \texttt{ //S2 superior}$ \\
            0                   & \quad $otherwise$                            & $\color{cyan} \texttt{ //equal}$ \\
        \end{array}
        \right.
    \end{array}
\end{equation}

A solution is always superior if it excludes less item area.
If two solutions include the same item area, the one with the highest total leftover value is superior. 
Therefore, leftover value can be viewed as a tiebreaker in the evaluation of solutions. 
Note that, despite the fact that the objective is to minimize the total bin area, this factor is not explicitly included in the cost function.
Total bin area is irrelevant so long as it remains below the limit. 
As mentioned in Section \ref{section:GDA}, the limit is lowered every time a new feasible solution is reached. 

\subsubsection{Late-Acceptance Hill-Climbing}\label{section:lahc}
GDRR employs the Late Acceptance Hill-Climbing (LAHC) metaheuristic, introduced by \citet{burke2017late}. 
This local search algorithm accepts non-improving moves when a candidate solution is better than the best solution a certain number of iterations ago. 
The pseudocode in Algorithm \ref{alg:GDRR} shows how LAHC is implemented.

The main argument for using LAHC is its scale independence.
Given that LAHC is insensitive to the scale of the objective function, the metaheuristic does not need to be tuned on an instance-specific basis. 
This is a major advantage over metaheuristics that make use of cooling schemes, such as Simulated Annealing.

Furthermore, since solutions are accepted based on their superiority with respect to previous solutions, no absolute difference in cost is needed. 
The only requirement is being able to compare two costs.
No delta value is necessary. 
This allows for the easy use of tiebreakers, such as the one outlined in Section \ref{section:solutionquality}, where the amount of included item area always dominates leftover value.

The final argument for using LAHC is that there is only a single parameter to tune: the history length ($\mathcal{L}_h$). This parameter defines how far back to look when evaluating a solution.

\begin{algorithm}[H]
		\caption{GDRR}\label{alg:GDRR}
		\begin{tabbing}
			\hspace*{\algorithmicindent} \textbf{Input:} \hspace*{2mm} \= A starting solution $S = \{C, E\}$ where $C$ is a set of cutting patterns\\
			\> and $E$ is a set of excluded items, bin area limit $\mathcal A_{lim}$,\\
			\> history length $\mathcal L_h$, the average number of nodes to remove $\mu$,\\
			\> and the current best solution $S_{best}$ \\
			\hspace*{\algorithmicindent} \textbf{Output:} The best (complete) solution $S_{best}$
		\end{tabbing}
		\begin{algorithmic}[1]
			\Function{gdrr}{$S, \mathcal A_{lim}, \mathcal L_h, \mu, S_{best}$}
			\ForAll{$k \in \{0,...,\mathcal L_h-1\} $}
				$S_k \gets S$
			\EndFor
			\State $i \gets 0$
			\State $S^* = \{C^*, E^*\} \gets S$ $\color{cyan} \texttt{ //local optimum (incomplete)}$
			\While{\text{no stopping criteria met}}
				\State $S' \gets $ \Call{ruin}{$S^*, \mathcal A_{lim}, \mu$} 
				\State $S' \gets $ \Call{recreate}{$S', \mathcal A_{lim}$} 
				\State $v \gets  i$ mod $\mathcal L_h$
				\If{$compare(S', S_v) \leq 0$ \textbf{or} $compare(S', S^{*}) \leq 0$} $\color{cyan} \texttt{ //solution accepted}$
					\State $S^* \gets S'$ 
					\If{$compare(S^*, S_v) < 0$} $\color{cyan} \texttt{ //fitness array updated}$
						\State $S_v \gets S^*$
					\EndIf
					\State $i \gets i + 1$
				\EndIf
				\If{$E^* = $ \O} $\color{cyan} \texttt{ //complete solution reached}$
					\State $S_{best} \gets S^*$
					\State $\mathcal A_{lim}' \gets \sum_{c \in C^*} \mathcal A_c$ 
					\State $S_{next} \gets \Call{ruin}{S_{best}, \mathcal A_{lim}', 0}$ 
					\State $S_{best} \gets \Call{gdrr}{S_{next}, \mathcal A_{lim}', \mathcal L_h, \mu, S_{best}}$
				\EndIf
			\EndWhile
			\State \Return $S_{best}$
			\EndFunction
		\end{algorithmic}
	\end{algorithm}

Algorithm \ref{alg:GDRR} provides a high-level overview of GDRR.
As mentioned in Section \ref{section:GDA}, the heuristic will continuously attempt to fit all items inside the bins while respecting the bin area limit constraint. 
This limit is lowered each time the heuristic is able to construct a feasible solution.

	The LAHC fitness array is initially filled entirely with the starting solution (line 2). This array keeps track of the last $\mathcal{L}_h$ best solutions.
	During each local search iteration, the solution is partially destroyed and rebuilt (lines 6-7) according to the procedures described in Sections \ref{section:ruin} and \ref{section:recreate}.
	Index $v$ corresponds to the index of the first element in the fitness array.
	Instead of shifting all elements of the fitness array every time a new solution is accepted, the head of the array is moved (line 8). 
	This is similar to how a circular buffer works. 
	In order for a modified solution ($S'$) to be accepted, it must either be superior or equal to the best solution $\mathcal{L}_h$ iterations ago ($S_v$) or improve upon the local optimum ($S^*$) (lines 9-10).
	The fitness array is only updated with improving values (line 11-12).
	The implementation of LAHC is analogous to that outlined in \citet{burke2017late}, but with one major exception.
	Counter $i$ is only incremented when a solution is accepted (line 13). 
	This is in contrast to Burke and Bykov's original LAHC, where the counter is always incremented.
	The main motivation behind this decision is to avoid the heuristic from becoming a pure hill-climbing algorithm in later stages and thus easily getting stuck in local optima.

	If a solution is obtained that has no excluded items (line 14), GDRR has reached its current goal.
	When this occurs, the best complete solution $S_{best}$ is updated (line 15), a new bin area limit $\mathcal{A}'_{lim}$ is calculated based on the total bin area of $S_{best}$ (line 16) and a starting solution for the next goal $S_{next}$ is created which complies with this new limit (line 17).
	Finally, GDRR is `restarted' for the next goal (line 18).
	Once a stopping criterion is met, the best complete solution is returned (line 19).

	In essence, the heuristic continuously strives towards feasibility with an iteratively lowering bin area limit.
	To initiate the optimization, Algorithm \ref{alg:GDRR} can be called with an empty $S$ (no cutting patterns and all items excluded) and with $\mathcal{A}_{lim}$ set to $\infty$. 
	There is no need for a separate algorithm to create an initial solution.

\subsection{Multithreading}\label{section:multithreading}
Multithreading can help to improve the performance of the heuristic and makes use of the capabilities of modern processors.
Each thread runs its own separate instance of GDRR, but a single global bin area limit is shared across all threads. 
This means that once one thread reaches a feasible solution, the limit is lowered across all threads.
Solutions themselves are, by contrast, not shared between threads. 
This is mainly to preserve diversity.

The multithreaded implementation is largely analogous to Algorithm \ref{alg:GDRR}. The only major difference is that the bin area limit is not only lowered when the heuristic reaches a feasible solution (lines 12-14), but rather whenever any thread reaches a feasible solution at the current limit.
It goes without saying that the best feasible solution among all threads is saved.

Threads are almost entirely independent and only influence each other indirectly. There are, for example, no explicit mechanisms in place to avoid duplicate searched areas.
Due to the vast search space and the heuristic containing a number of stochastic elements (e.g. blinks), there is no real need for such mechanisms.

Multithreading is primarily used to maintain diversity, not necessarily to accelerate execution speed.
The use of multithreading is particularly beneficial for instances with many different bins.
At later stages of the optimization, solutions are often too fixed to significantly change the combination of used bins.
Multithreading helps overcome this issue since different threads can have solutions that are close in terms of quality, but consist of very different combinations of bins.
This level of diversity would be difficult to achieve with only a single thread, as it would require very extreme ruin procedures.

\section{Computational Results} \label{section:computationalresults}
To evaluate the performance of GDRR, the algorithm was tested on a number of different benchmarks from the literature. These benchmarks contain datasets with both identical and heterogeneous sets of bins.
\citet{ORdatasets} provides a central repository for a large number of cutting and packing datasets from the literature. All datasets included in the repository are formatted in the same structure, which saves a lot of time when testing across a number of different benchmarks.

Section \ref{section:parameters} defines how the parameters of GDRR were configured for the computational experiments. The performance of the algorithm is then evaluated in Sections \ref{section:2BPOGresults}-\ref{section:2VSBPRGresults}.
Finally, Sections \ref{section:impactmultithreading} and \ref{section:impactruntime} analyze the impact of multithreading and calculation time on the performance of GDRR.
All experiments were conducted on an Intel Xeon Gold 6240 CPU (2.6GHz) with 8GB of RAM. For the comparisons in Sections \ref{section:2BPOGresults}-\ref{section:2VSBPRGresults}, GDRR was configured to use 8 threads and 600 seconds of runtime.
Solutions for all datasets tested in the comparison are included as an online supplement of this paper.

\subsection{Parameters}	\label{section:parameters}
GDRR has a number of parameters which must be configured.
The tuning of these parameters was conducted using the dataset from \citet{ortmann2010new}. 
This dataset was chosen because it contains a large variety of instances and has variable-sized bins. Heterogeneous bins enable gradual improvements in solution quality to be visible. 
All parameters are tuned to achieve best results when optimizing for 600 seconds and 8 threads.
Parameters are fixed and varied one by one to examine their individual effect on the behavior of the algorithm. 
However, the history length ($\mathcal{L}_h$) and the average number of removed nodes ($\mu$) were tuned as a pair since they are so closely related.
The parameters listed below are used throughout the experiments:

\begin{itemize}
	\item \textbf{Leftover valuation power: $\alpha$}\\
	      In Section \ref{section:insertionoptions}, parameter $\alpha$ was defined as the power with which to multiply the area of a leftover node in order to derive its value. The leftover value is not only used to evaluate insertion options, but also as a tiebreaker when determining a solution's quality (Section \ref{section:solutionquality}).	
	      A greater $\alpha$ will favor few, but large, leftover areas. When $\alpha$ is smaller, many medium-sized leftover pieces are not penalized as much.
	      Experiments show that the value of $\alpha$ has a minimal impact so long as it is greater than 1. A value of $1.2$ resulted in the best overall performance.
	      \begin{itemize}
		      \item $\alpha=1.2$
	      \end{itemize}
	      
	\item \textbf{History Length: $\mathcal{L}_h$} \\
	      $\mathcal{L}_h$ determines the length of the history queue in LAHC.
	      The best value for $\mathcal{L}_h$ depends largely on the size of the instance. The larger an instance, the shorter the optimal history length. This is because larger instances result in less ruin and recreate iterations per second and generally have more diverse neighborhoods. Preliminary experiments resulted in the following rules of thumb:  
	      \begin{itemize}
		      \item $0 < \#items \leq 100$: $\mathcal{L}_h = 2000$
		      \item $100 < \#items \leq 300$: $\mathcal{L}_h = 1000$
		      \item $300 < \#items \leq 500$: $\mathcal{L}_h = 500$
	      \end{itemize}
	\item \textbf{Average number of removed nodes: $\mu$} \\
	      As with $\mathcal{L}_h$, the optimal value of $\mu$ depends on the instance size. Ruining the solution less significantly results in more iterations per second, which is beneficial for larger instances. Preliminary experiments showed that the following values for $\mu$ are a suitable choice:
	      
	      \begin{itemize}
		      \item $0 < \#items \leq 100$: $\mu = 8$
		      \item $100 < \#items \leq 300$: $\mu = 6$
		      \item $300 < \#items \leq 500$: $\mu = 4$
	      \end{itemize}
	      
	\item \textbf{Blink rate : $\beta$} \\
	      Across the dataset, a blinking chance of 5\% was deemed to be a good balance between greediness and randomness.
	      \begin{itemize}
		      \item $\beta = 0.05$
	      \end{itemize}
	      
\end{itemize}

\subsection{Results for 2BP\textbar O\textbar G} \label{section:2BPOGresults}
To compare the performance of GDRR for the 2BP with guillotine cuts and no rotation of items, the well-known instances from \citet{berkey1987two} (classes 1-6) and \citet{lodi1999heuristic} (classes 7-10) are used. Each class has ten instances of 20, 40, 60, 80 and 100 items. As a result, the dataset contains a total of 500 instances.

A comparison between GDRR and other heuristics in the literature is presented in Table \ref{tab:2dbpog_bmwv}.
The sum of bins for each instance category is displayed. As \citet{Cui2018} note, the majority of instances from this dataset have already been solved to optimality. Therefore, the differences between algorithms might seem small. 
The comparison includes a partial enumeration heuristic (ENH) by \citet{Lodi2017}, a constructive heuristic (CHBP) by \citet{Charalambous2011}, a hybrid heuristic algorithm (HHA) by \citet{Hong2014}, the critical fit insertion heuristic (CFIH) by \citet{fleszar2013three} and the hybrid heuristic with triple block patterns (HHTB) by \citet{Cui2018}. For all of these heuristics, the best results reported in each paper are presented in the comparison.

The proposed heuristic achieves the best scores in each category. However, it should be noted that, with the exception of ENH\cite{Lodi2017}, GDRR has longer runtimes than most of the other algorithms. Since many of these algorithms are based heavily on constructive approaches their runtimes are around or under the 1-minute mark.

\begin{table}[H]
	\centering
	\resizebox{0.85\columnwidth}{!}{

		\begin{tabular}{llllllll} 
			\hline
			Class               & Number of items & GDRR            & ENH \cite{Lodi2017}         & CHBP \cite{Charalambous2011}        & HHA  \cite{Hong2014}         & CFIH \cite{fleszar2013three}       & HHTB \cite{Cui2018}          \\ 
			\hline
			\multirow{6}{*}{1}  & 20              & \textbf{71}   & \textbf{71}   & \textbf{71}  & \textbf{71}   & \textbf{71}   & \textbf{71}    \\
			& 40              & \textbf{134}  & \textbf{134}  & \textbf{134} & \textbf{134}  & 135           & \textbf{134}   \\
			& 60              & \textbf{200}  & \textbf{200}  & 201          & 201           & 201           & \textbf{200}   \\
			& 80              & \textbf{275}  & \textbf{275}  & \textbf{275} & \textbf{275}  & \textbf{275}  & \textbf{275}   \\
			& 100             & \textbf{317}  & \textbf{317}  & 321          & 320           & 322           & \textbf{317}   \\
			& All             & \textbf{997}  & \textbf{997}  & 1002         & 1001          & 1004          & \textbf{997}   \\ 
			\hline
			\multirow{6}{*}{2}  & 20              & \textbf{10}   & \textbf{10}   & \textbf{10}  & \textbf{10}   & \textbf{10}   & \textbf{10}    \\
			& 40              & \textbf{19}   & 20            & 20           & \textbf{19}   & 20            & 20             \\
			& 60              & \textbf{25}   & \textbf{25}   & 26           & \textbf{25}   & 27            & \textbf{25}    \\
			& 80              & \textbf{31}   & 32            & 33           & \textbf{31}   & 32            & \textbf{31}    \\
			& 100             & \textbf{39}   & \textbf{39}   & \textbf{39}  & \textbf{39}   & 40            & \textbf{39}    \\
			& All             & \textbf{124}  & 126           & 128          & \textbf{124}  & 129           & 125            \\ 
			\hline
			\multirow{6}{*}{3}  & 20              & \textbf{52}   & 54            & \textbf{52}  & \textbf{52}   & 53            & \textbf{52}    \\
			& 40              & \textbf{94}   & 96            & 97           & \textbf{94}   & 96            & \textbf{94}    \\
			& 60              & \textbf{139}  & 140           & 140          & 140           & 141           & \textbf{139}   \\
			& 80              & \textbf{189}  & 190           & 196          & 192           & 195           & \textbf{189}   \\
			& 100             & \textbf{223}  & 225           & 230          & 228           & 226           & \textbf{223}   \\
			& All             & \textbf{697}  & 705           & 715          & 706           & 711           & \textbf{697}   \\ 
			\hline
			\multirow{6}{*}{4}  & 20              & \textbf{10}   & \textbf{10}   & \textbf{10}  & \textbf{10}   & \textbf{10}   & \textbf{10}    \\
			& 40              & \textbf{19}   & \textbf{19}   & \textbf{19}  & \textbf{19}   & \textbf{19}   & \textbf{19}    \\
			& 60              & \textbf{24}   & 25            & 25           & 25            & 25            & 25             \\
			& 80              & \textbf{31}   & 33            & 33           & 32            & 33            & \textbf{31}    \\
			& 100             & \textbf{37}   & 39            & 39           & 38            & 39            & \textbf{37}    \\
			& All             & \textbf{121}  & 126           & 126          & 124           & 126           & 122            \\ 
			\hline
			\multirow{6}{*}{5}  & 20              & \textbf{65}   & 66            & \textbf{65}  & \textbf{65}   & 66            & \textbf{65}    \\
			& 40              & \textbf{119}  & \textbf{119}  & 121          & \textbf{119}  & 120           & \textbf{119}   \\
			& 60              & \textbf{180}  & 181           & 183          & 181           & 182           & \textbf{180}   \\
			& 80              & \textbf{247}  & \textbf{247}  & \textbf{247} & \textbf{247}  & 248           & \textbf{247}   \\
			& 100             & \textbf{282}  & 286           & 288          & 287           & 290           & \textbf{282}   \\
			& All             & \textbf{893}  & 899           & 904          & 899           & 906           & \textbf{893}   \\ 
			\hline
			\multirow{6}{*}{6}  & 20              & \textbf{10}   & \textbf{10}   & \textbf{10}  & \textbf{10}   & \textbf{10}   & \textbf{10}    \\
			& 40              & \textbf{17}   & 19            & 19           & 18            & 19            & \textbf{17}    \\
			& 60              & \textbf{21}   & 22            & 22           & 22            & 22            & \textbf{21}    \\
			& 80              & \textbf{30}   & \textbf{30}   & \textbf{30}  & \textbf{30}   & \textbf{30}   & \textbf{30}    \\
			& 100             & \textbf{32}   & 35            & 34           & 35            & 34            & \textbf{32}    \\
			& All             & \textbf{110}  & 116           & 115          & 115           & 115           & \textbf{110}   \\ 
			\hline
			\multirow{6}{*}{7}  & 20              & \textbf{55}   & \textbf{55}   & \textbf{55}  & \textbf{55}   & 56            & \textbf{55}    \\
			& 40              & \textbf{111}  & 113           & 112          & \textbf{111}  & 115           & \textbf{111}   \\
			& 60              & \textbf{158}  & 159           & 160          & 159           & 161           & \textbf{158}   \\
			& 80              & \textbf{230}  & 232           & 233          & 232           & 232           & 231            \\
			& 100             & \textbf{271}  & 275           & 275          & 273           & 274           & \textbf{271}   \\
			& All             & \textbf{825}  & 834           & 835          & 830           & 838           & 826            \\ 
			\hline
			\multirow{6}{*}{8}  & 20              & \textbf{58}   & \textbf{58}   & \textbf{58}  & \textbf{58}   & 60            & \textbf{58}    \\
			& 40              & \textbf{113}  & \textbf{113}  & 114          & \textbf{113}  & 116           & \textbf{113}   \\
			& 60              & \textbf{161}  & 162           & 163          & 162           & 165           & \textbf{161}   \\
			& 80              & \textbf{224}  & 226           & 226          & 225           & 227           & \textbf{224}   \\
			& 100             & \textbf{277}  & 280           & 279          & 279           & 281           & \textbf{277}   \\
			& All             & \textbf{833}  & 839           & 840          & 837           & 849           & \textbf{833}   \\ 
			\hline
			\multirow{6}{*}{9}  & 20              & \textbf{143}  & \textbf{143}  & \textbf{143} & \textbf{143}  & \textbf{143}  & \textbf{143}   \\
			& 40              & \textbf{278}  & \textbf{278}  & 279          & \textbf{278}  & \textbf{278}  & \textbf{278}   \\
			& 60              & \textbf{437}  & \textbf{437}  & 438          & \textbf{437}  & \textbf{437}  & \textbf{437}   \\
			& 80              & \textbf{577}  & \textbf{577}  & \textbf{577} & \textbf{577}  & \textbf{577}  & \textbf{577}   \\
			& 100             & \textbf{695}  & \textbf{695}  & \textbf{695} & \textbf{695}  & \textbf{695}  & \textbf{695}   \\
			& All             & \textbf{2130} & \textbf{2130} & 2132         & \textbf{2130} & \textbf{2130} & \textbf{2130}  \\ 
			\hline
			\multirow{6}{*}{10} & 20              & \textbf{42}   & 44            & 44           & 43            & 43            & 43             \\
			& 40              & \textbf{74}   & \textbf{74}   & \textbf{74}  & \textbf{74}   & 75            & \textbf{74}    \\
			& 60              & \textbf{101}  & 102           & 103          & 102           & 104           & \textbf{101}   \\
			& 80              & \textbf{128}  & 130           & 130          & 130           & 132           & \textbf{128}   \\
			& 100             & \textbf{158}  & 159           & 163          & 160           & 163           & 159            \\
			& All             & \textbf{503}  & 509           & 514          & 509           & 517           & 505            \\ 
			\hline
			Total bins:              &                 & \textbf{7233} & 7281          & 7311         & 7275          & 7325          & 7238           \\
			\hline
		\end{tabular}
	}
	\caption{Results for the 2DBP$\vert$O$\vert$G benchmark from \citet{berkey1987two} and \citet{lodi1999heuristic}}
	\label{tab:2dbpog_bmwv}
\end{table}

\begin{table}[H]
	\centering
	\resizebox{0.85\columnwidth}{!}{
		\begin{tabular}{llllllllll} 
			\cline{1-8}
			Class               & Number of items & GDRR            & CHBP \cite{Charalambous2011}         & A-B \cite{Polyakovsky2009}         & SVCRG   \cite{Cui2015}      & CFIH   \cite{fleszar2013three}        & HHTB  \cite{Cui2018} &  &   \\ 
			\cline{1-8}
			\multirow{6}{*}{1}  & 20              & \textbf{66}   & \textbf{66}   & \textbf{66}   & \textbf{66}   & \textbf{66}   & \textbf{66}   &  &   \\
			& 40              & \textbf{128}  & 129           & 129           & \textbf{128}  & 129           & \textbf{128}  &  &   \\
			& 60              & \textbf{195}  & \textbf{195}  & \textbf{195}  & \textbf{195}  & \textbf{195}  & \textbf{195}  &  &   \\
			& 80              & \textbf{270}  & 271           & \textbf{270}  & \textbf{270}  & 271           & \textbf{270}  &  &   \\
			& 100             & \textbf{313}  & 314           & 314           & 314           & 314           & \textbf{313}  &  &   \\
			& All             & \textbf{972}  & 975           & 974           & 973           & 975           & \textbf{972}  &  &   \\ 
			\cline{1-8}
			\multirow{6}{*}{2}  & 20              & \textbf{10}   & \textbf{10}   & \textbf{10}   & \textbf{10}   & \textbf{10}   & \textbf{10}   &  &   \\
			& 40              & \textbf{19}   & \textbf{19}   & \textbf{19}   & 20            & \textbf{19}   & 20            &  &   \\
			& 60              & \textbf{25}   & \textbf{25}   & \textbf{25}   & 29            & \textbf{25}   & \textbf{25}   &  &   \\
			& 80              & \textbf{31}   & \textbf{31}   & \textbf{31}   & 33            & 32            & \textbf{31}   &  &   \\
			& 100             & \textbf{39}   & \textbf{39}   & \textbf{39}   & 40            & \textbf{39}   & \textbf{39}   &  &   \\
			& All             & \textbf{124}  & \textbf{124}  & \textbf{124}  & 132           & 125           & 125           &  &   \\ 
			\cline{1-8}
			\multirow{6}{*}{3}  & 20              & \textbf{47}   & 48            & 49            & \textbf{47}   & 48            & \textbf{47}   &  &   \\
			& 40              & \textbf{92}   & 94            & 94            & \textbf{92}   & 94            & \textbf{92}   &  &   \\
			& 60              & \textbf{134}  & 136           & 137           & 135           & 136           & \textbf{134}  &  &   \\
			& 80              & \textbf{183}  & 186           & 188           & 184           & 185           & \textbf{183}  &  &   \\
			& 100             & \textbf{219}  & 223           & 223           & 223           & 223           & \textbf{219}  &  &   \\
			& All             & \textbf{675}  & 687           & 691           & 681           & 686           & \textbf{675}  &  &   \\ 
			\cline{1-8}
			\multirow{6}{*}{4}  & 20              & \textbf{10}   & \textbf{10}   & \textbf{10}   & \textbf{10}   & \textbf{10}   & \textbf{10}   &  &   \\
			& 40              & \textbf{19}   & \textbf{19}   & \textbf{19}   & \textbf{19}   & \textbf{19}   & \textbf{19}   &  &   \\
			& 60              & \textbf{23}   & 25            & 25            & 25            & 25            & 25            &  &   \\
			& 80              & \textbf{30}   & 33            & 31            & 31            & 33            & 31            &  &   \\
			& 100             & \textbf{37}   & 38            & \textbf{37}   & 38            & 38            & \textbf{37}   &  &   \\
			& All             & \textbf{119}  & 125           & 122           & 123           & 125           & 122           &  &   \\ 
			\cline{1-8}
			\multirow{6}{*}{5}  & 20              & \textbf{59}   & \textbf{59}   & 60            & \textbf{59}   & \textbf{59}   & \textbf{59}   &  &   \\
			& 40              & \textbf{114}  & 115           & 115           & \textbf{114}  & 117           & \textbf{114}  &  &   \\
			& 60              & \textbf{173}  & 176           & 176           & 174           & 175           & \textbf{173}  &  &   \\
			& 80              & 239           & 240           & 243           & 240           & 241           & \textbf{238}  &  &   \\
			& 100             & \textbf{277}  & 282           & 284           & 281           & 281           & \textbf{277}  &  &   \\
			& All             & 862           & 872           & 878           & 868           & 873           & \textbf{861}  &  &   \\ 
			\cline{1-8}
			\multirow{6}{*}{6}  & 20              & \textbf{10}   & \textbf{10}   & \textbf{10}   & \textbf{10}   & \textbf{10}   & \textbf{10}   &  &   \\
			& 40              & \textbf{16}   & 18            & 18            & 19            & 18            & 18            &  &   \\
			& 60              & \textbf{21}   & \textbf{21}   & \textbf{21}   & \textbf{21}   & \textbf{21}   & \textbf{21}   &  &   \\
			& 80              & \textbf{30}   & \textbf{30}   & \textbf{30}   & \textbf{30}   & \textbf{30}   & \textbf{30}   &  &   \\
			& 100             & \textbf{32}   & 34            & 34            & 34            & 34            & \textbf{32}   &  &   \\
			& All             & \textbf{109}  & 113           & 113           & 114           & 113           & 111           &  &   \\ 
			\cline{1-8}
			\multirow{6}{*}{7}  & 20              & \textbf{52}   & \textbf{52}   & \textbf{52}   & \textbf{52}   & \textbf{52}   & \textbf{52}   &  &   \\
			& 40              & 102           & 104           & 103           & 103           & 106           & \textbf{101}  &  &   \\
			& 60              & \textbf{145}  & 148           & 148           & \textbf{145}  & 151           & \textbf{145}  &  &   \\
			& 80              & 208           & 211           & 211           & 210           & 214           & \textbf{207}  &  &   \\
			& 100             & 249           & 255           & 252           & 251           & 258           & \textbf{248}  &  &   \\
			& All             & 756           & 770           & 766           & 761           & 781           & \textbf{753}  &  &   \\ 
			\cline{1-8}
			\multirow{6}{*}{8}  & 20              & \textbf{53}   & \textbf{53}   & \textbf{53}   & \textbf{53}   & \textbf{53}   & \textbf{53}   &  &   \\
			& 40              & 104           & 105           & 104           & 104           & 105           & \textbf{103}  &  &   \\
			& 60              & \textbf{146}  & 150           & 150           & 147           & 152           & \textbf{146}  &  &   \\
			& 80              & 204           & 210           & 209           & 207           & 211           & \textbf{203}  &  &   \\
			& 100             & \textbf{252}  & 258           & 256           & 254           & 258           & \textbf{252}  &  &   \\
			& All             & 759           & 776           & 772           & 765           & 779           & \textbf{757}  &  &   \\ 
			\cline{1-8}
			\multirow{6}{*}{9}  & 20              & \textbf{143}  & \textbf{143}  & \textbf{143}  & \textbf{143}  & \textbf{143}  & \textbf{143}  &  &   \\
			& 40              & \textbf{275}  & \textbf{275}  & \textbf{275}  & \textbf{275}  & \textbf{275}  & \textbf{275}  &  &   \\
			& 60              & \textbf{435}  & \textbf{435}  & \textbf{435}  & \textbf{435}  & \textbf{435}  & \textbf{435}  &  &   \\
			& 80              & \textbf{573}  & \textbf{573}  & \textbf{573}  & \textbf{573}  & \textbf{573}  & \textbf{573}  &  &   \\
			& 100             & \textbf{693}  & \textbf{693}  & \textbf{693}  & \textbf{693}  & \textbf{693}  & \textbf{693}  &  &   \\
			& All             & \textbf{2119} & \textbf{2119} & \textbf{2119} & \textbf{2119} & \textbf{2119} & \textbf{2119} &  &   \\ 
			\cline{1-8}
			\multirow{6}{*}{10} & 20              & \textbf{41}   & \textbf{41}   & \textbf{41}   & \textbf{41}   & 42            & \textbf{41}   &  &   \\
			& 40              & \textbf{72}   & 73            & 73            & 73            & 73            & 73            &  &   \\
			& 60              & \textbf{99}   & 100           & 100           & 100           & 101           & \textbf{99}   &  &   \\
			& 80              & \textbf{124}  & 130           & 129           & 126           & 129           & 126           &  &   \\
			& 100             & \textbf{155}  & 159           & 161           & 159           & 159           & \textbf{155}  &  &   \\
			& All             & \textbf{491}  & 503           & 504           & 499           & 504           & 494           &  &   \\ 
			\cline{1-8}
			Total bins:              &                 & \textbf{6986} & 7064          & 7063          & 7035          & 7080          & 6989          &  &   \\ 
			\cline{1-8}
			&                 &               &               &               &               &               &               &  &   \\
			&                 &               &               &               &               &               &               &  &  
		\end{tabular}
	}
	\caption{Results for the 2DBP$\vert$R$\vert$G benchmark from \citet{berkey1987two} and \citet{lodi1999heuristic}}
	\label{tab:2dbprg_bmwv}
\end{table}

\subsection{Results for 2BP\textbar R\textbar G} \label{section:2BPRGresults}
To compare the performance of the heuristic for the 2BP with guillotine cuts and 90$\degree$ rotation of items, once again the dataset by \citet{berkey1987two} and \citet{lodi1999heuristic} is used. 
In addition to some of the heuristics from Section \ref{section:2BPOGresults}, this comparison also includes an agent-based approach (A-B) by \citet{Polyakovsky2009} and the sequential value correcting heuristic (SVCRG) by \citet{Cui2015}.

Unlike in Section \ref{section:2BPOGresults}, GDRR does not always come out on top here. On average the proposed heuristic achieves the best quality solutions, but there are some classes where HHTB is superior.

\subsection{Results for 2VSBP\textbar O\textbar G}\label{section:2VSBPOGresults}
The performance on instances with a heterogeneous set of bins is tested using three well-known benchmarks.  
In this comparison, the average utilization is compared across different algorithms.
Utilization $\gamma$ corresponds to the percentage of area of the bins that are packed with items. A utilization of 95\%, for example, corresponds to 5\% wasted bin area. 
\begin{equation*}
	\gamma = \frac{\text{total item area}}{\text{total bin area}} \cdot 100\%
\end{equation*}

All the results from other heuristics in the tables are taken from the comparison performed by \citet{Hong2014}.
Table \ref{tab:2vsdbpog_m} provides a comparison for the instances created by \citet{hopper2002problem} which consist of three problem categories, each containing five instances. The second dataset was introduced by \citet{pisinger2005two} and is divided into ten classes, the results for which are summarized in Table \ref{tab:2vsdbpog_ps}.
Note that this dataset does not define limits on the quantity of bins of each type.
The last dataset, introduced by \citet{ortmann2010new}, contains instances based on nice (relatively normal) and pathologically (unusual, more extreme) sized items. These instances range from 25-500 items and 2-6 different bin sizes. Table \ref{tab:2vsdbpog_nicepath} provides the results compared to other heuristics. For all of these datasets and classes, GDRR always achieves the best solutions in terms of quality.

\begin{table}[h]
\centering
\resizebox{1\columnwidth}{!}{
\begin{tabular}{lllllllllll} 
	\hline
	Class   & GDRR & FFDH\cite{coffman1980performance}  & BFDH\cite{coffman1990average}  & JOIN\cite{martello2003exact}  & FCOG\cite{lodi1999heuristic} & BFDH*\cite{bortfeldt2006genetic} & SAS\cite{ntene2009survey}   & SASm\cite{ortmann2010new}  & BFS\cite{ortmann2010new}  & HHA\cite{Hong2014}    \\ 
	\hline
	M1      & \textbf{98.4}  & 93.5  & 93.5  & 86.8  & 94.9  & 94.9  & 83.5  & 91.6  & 95.5  & \textbf{98.4}  \\
	M2      & \textbf{97.2}  & 87.5  & 88.8  & 82.8  & 89.6  & 89.6  & 81.7  & 88.0  & 90.0  & 95.6           \\
	M3      & \textbf{98.0}  & 92.0  & 92.6  & 85.5  & 93.6  & 93.6  & 86.8  & 91.8  & 94.9  & 97.4           \\ 
	\hline
	Average utilization [\%]: & \textbf{97.85} & 91.00 & 91.63 & 85.03 & 92.70 & 92.70 & 84.00 & 90.47 & 93.47 & 97.13          \\
	\hline
\end{tabular}

}
\caption{Results for the 2VSBP$\vert$O$\vert$G benchmark from \citet{hopper2002problem}}
	\label{tab:2vsdbpog_m}
\end{table}

\begin{table}[h]
	\centering
	\resizebox{1\columnwidth}{!}{
\begin{tabular}{lllllllllll}
	\hline
	Class   & GDRR & FFDH\cite{coffman1980performance}  & BFDH\cite{coffman1990average}  & JOIN\cite{martello2003exact}  & FCOG\cite{lodi1999heuristic} & BFDH*\cite{bortfeldt2006genetic} & SAS\cite{ntene2009survey}   & SASm\cite{ortmann2010new}  & BFS\cite{ortmann2010new}  & HHA\cite{Hong2014}    \\ 
	\hline
	I                            & \textbf{94.1}  & 86.3  & 86.6  & 83.0  & 87.3  & 87.4  & 79.4 & 86.3  & 88.6  & 91.5  \\
	II                           & \textbf{96.5}  & 83.7  & 83.7  & 80.9  & 85.2  & 85.0  & 80.1 & 82.2  & 85.1  & 96.3  \\
	III                          & \textbf{91.5}  & 81.1  & 81.7  & 76.7  & 81.9  & 81.9  & 69.6 & 75.7  & 82.2  & 86.6  \\
	IV                           & \textbf{93.5}  & 80.0  & 80.0  & 79.1  & 82.7  & 82.1  & 76.0 & 78.0  & 82.0  & 91.7  \\
	V                            & \textbf{89.3}  & 80.4  & 81.1  & 77.6  & 81.3  & 81.2  & 72.6 & 78.3  & 81.4  & 84.6  \\
	VI                           & \textbf{92.7}  & 79.1  & 79.3  & 78.3  & 80.7  & 80.5  & 76.1 & 77.1  & 79.5  & 90.2  \\
	VII                          & \textbf{90.1}  & 79.9  & 80.2  & 79.6  & 80.8  & 80.6  & 74.0 & 80.4  & 80.9  & 86.9  \\
	VIII                         & \textbf{89.4}  & 80.7  & 81.1  & 74.2  & 81.3  & 81.2  & 76.4 & 79.5  & 81.6  & 85.9  \\
	IX                           & \textbf{75.6}  & 72.8  & 72.6  & 72.1  & 72.6  & 72.8  & 71.6 & 72.9  & 73.0  & 74.3  \\
	X                            & \textbf{93.0}  & 83.3  & 83.8  & 79.3  & 85.4  & 84.6  & 73.2 & 79.4  & 85.5  & 90.3  \\ \hline
	Average	utilization [\%]:					 & \textbf{90.66} & 80.73 & 81.01 & 78.08 & 81.92 & 81.73 & 74.90 & 78.98 & 81.98 & 87.83 \\ \hline
\end{tabular}
}
\caption{Results for the 2VSBP$\vert$O$\vert$G benchmark from \citet{pisinger2005two}}
	\label{tab:2vsdbpog_ps}
\end{table}

\begin{table}[h]
	\centering
		\resizebox{1\columnwidth}{!}{
\begin{tabular}{lllllllllll}
\hline
Class   & GDRR & FFDH\cite{coffman1980performance}  & BFDH\cite{coffman1990average}  & JOIN\cite{martello2003exact}  & FCOG\cite{lodi1999heuristic} & BFDH*\cite{bortfeldt2006genetic} & SAS\cite{ntene2009survey}   & SASm\cite{ortmann2010new}  & BFS\cite{ortmann2010new}  & HHA\cite{Hong2014}    \\ 
\hline
Nice25i                      & \textbf{99.8}  & 73.9  & 73.6  & 70.6  & 73.9  & 73.6  & 68.3  & 71.8  & 73.6  & 94.3  \\
Nice50i                      & \textbf{93.5}  & 76.7  & 76.7  & 73.3  & 77.9  & 77.7  & 70.8  & 73.1  & 77.8  & 89.2  \\
Nice100i                     & \textbf{91.8}  & 79.4  & 79.4  & 77.5  & 79.9  & 79.4  & 75.7  & 76.3  & 79.4  & 88.3  \\
Nice200i                     & \textbf{92.6}  & 82.0  & 82.0  & 81.5  & 84.5  & 84.5  & 78.5  & 79.4  & 84.5  & 91.0  \\
Nice300i                     & \textbf{93.3}  & 85.8  & 85.8  & 83.3  & 86.0  & 86.5  & 80.2  & 81.7  & 86.8  & 92.0  \\
Nice400i                     & \textbf{94.3}  & 85.1  & 85.1  & 82.7  & 86.6  & 85.7  & 80.2  & 81.0  & 85.7  & 92.8  \\
Nice500i                     & \textbf{94.1}  & 87.2  & 87.2  & 84.8  & 87.7  & 87.7  & 81.8  & 83.8  & 87.7  & 93.1  \\ \hline
Nice                         & \textbf{94.04} & 81.44 & 81.40 & 79.10 & 82.35 & 82.15 & 76.50 & 78.15 & 82.21 & 91.52 \\ \hline
Path25i                      & \textbf{100}   & 76.3  & 76.3  & 73.9  & 77.9  & 77.6  & 72.2  & 74.4  & 78.3  & 97.2  \\
Path50i                      & \textbf{98.1}  & 76.4  & 78.6  & 74.2  & 81.6  & 79.4  & 72.4  & 75.6  & 81.9  & 94.9  \\
Path100i                     & \textbf{95.5}  & 79.7  & 79.7  & 77.8  & 83.2  & 81.3  & 72.4  & 75.8  & 83.5  & 93.5  \\
Path200i                     & \textbf{94.6}  & 84.1  & 84.0  & 81.8  & 88.0  & 85.9  & 77.7  & 79.0  & 87.5  & 93.3  \\
Path300i                     & \textbf{95.5}  & 82.9  & 82.9  & 82.7  & 87.0  & 86.0  & 81.4  & 81.7  & 87.3  & 94.9  \\
Path400i                     & \textbf{95.4}  & 82.7  & 82.7  & 82.3  & 89.6  & 87.0  & 79.9  & 80.1  & 88.5  & 94.8  \\
Path500i                     & \textbf{96.6}  & 82.9  & 82.9  & 81.2  & 88.7  & 86.4  & 81.3  & 82.4  & 86.7  & 94.0  \\ \hline
Path                         & \textbf{96.43} & 80.71 & 81.01 & 79.13 & 85.14 & 83.37 & 76.76 & 78.43 & 84.81 & 94.66 \\ \hline
Average utilization [\%]: 				     & \textbf{95.24} & 81.22 & 81.35 & 79.26 & 83.89 & 82.91 & 76.78 & 78.43 & 83.66 & 93.10 \\ \hline
\end{tabular}
}
\caption{Results for the 2VSBP$\vert$O$\vert$G benchmark from \citet{ortmann2010new}}
	\label{tab:2vsdbpog_nicepath}
\end{table}

\subsection{Results for 2VSBP\textbar R\textbar G}\label{section:2VSBPRGresults}
To the best of our knowledge, there are currently no algorithms available in the literature with which results for the 2BP variant with 90$\degree$ rotation and heterogeneous sets of bins can be compared. Since GDRR supports this variant, experiments were run on the same three datasets in Section \ref{section:2VSBPOGresults}. The solutions are included in the online supplement to facilitate comparisons in future research.

\subsection{Impact of multiple threads} \label{section:impactmultithreading}
\begin{figure}[h]
	\centering
	\begin{tikzpicture}
	\begin{axis}[
	xlabel={Number of threads},
	ylabel={Average Utilization [\%]},
	xmin=0, xmax=18,
	ymin=97, ymax=98.25,
	width=0.7\linewidth,
	height=.2\paperheight,
	%xmode=log,
	log ticks with fixed point,
	xtick={1,2,4,8,16},
	ytick={97,97.25,97.5,97.75,98,98.25},
	legend pos=north west,
	ymajorgrids=true,
	xmajorgrids=true,
	grid style=dotted,
	]
	
	\addplot[
	smooth,
	color=black,
	mark=*,
	]
	coordinates {
		(1,97.20314362)
		(2,97.47441202)
		(4,97.75552028)
		(8,97.91065277)
		(16,97.98422629)
	};
	
	\end{axis}
	\end{tikzpicture}
	\caption{Impact of multithreading on solution quality (600s of runtime)}
	\label{plot:multithreading}
\end{figure}
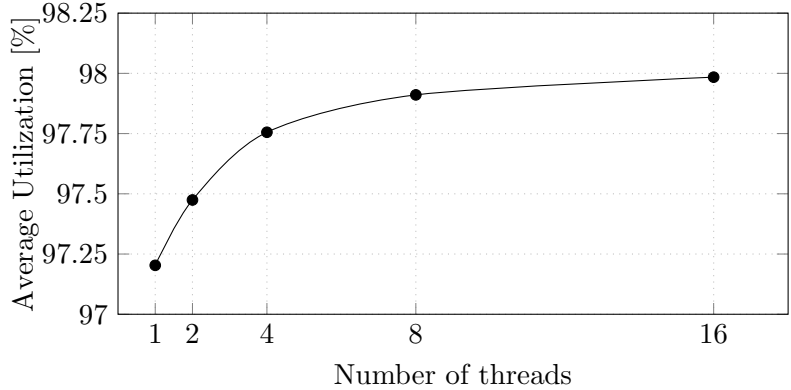

To test the impact of multithreading on the solution quality, experiments were run with a subset of instances from \citet{ortmann2010new}. 
From each class, the first instance with 6 different bins was selected. 
This decision was made because a greater variety of bins allows for more gradual differences in solution quality to be visible. 
The results are shown in Figure \ref{plot:multithreading}. 
As expected, the first few extra threads result in a significant performance improvement. 
Increasing the number of threads from one to four, for example, results in an improvement of 0.5\% in average utilization. However, adding additional threads quickly leads to diminishing returns.

\subsection{Impact of calculation time} \label{section:impactruntime}
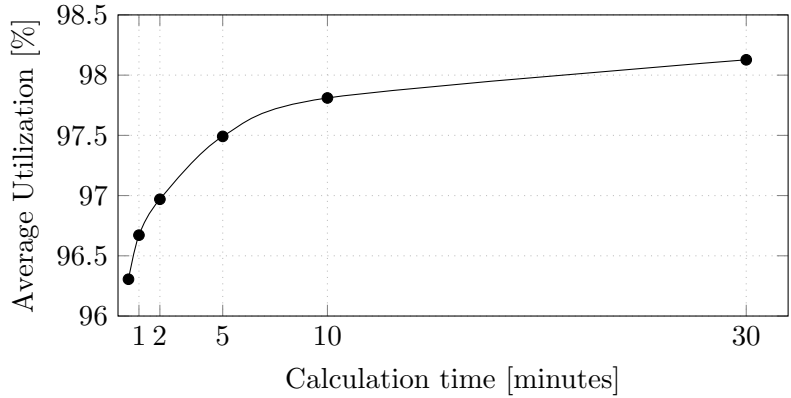
\begin{figure}[h]
	\centering
	\begin{tikzpicture}
	\begin{axis}[
	xlabel={Calculation time [minutes]},
	ylabel={Average Utilization [\%]},
	xmin=0, xmax=32,
	ymin=96, ymax=98.5,
	width=0.7\linewidth,
	height=.2\paperheight,
	%xmode=log,
	log ticks with fixed point,
	xtick={1, 2, 5, 10, 30},
	ytick={96,96.5,97,97.5,98,98.5,99},
	legend pos=north west,
	ymajorgrids=true,
	xmajorgrids=true,
	grid style=dotted,
	]
	
	\addplot[
	smooth,
	color=black,
	mark=*,
	]
	coordinates {
		(0.5,96.305601)
		(1,96.6714777)
		(2,96.96951729)
		(5,97.49141897)
		(10,97.81046666)
		(30,98.12731646)
	};
	
	\end{axis}
	\end{tikzpicture}
	\caption{Impact of calculation time on solution quality (8 threads)}
	\label{plot:runtime}
\end{figure}

The impact of available calculation time on solution quality was tested using the same subset of instances detailed in Section \ref{section:impactmultithreading}. 
Figure \ref{plot:runtime} shows how the heuristic scales with respect to calculation time.
Despite improvements slowing down, there are still gains to be had, even after 30 minutes. This is most likely because the history length ($\mathcal{L}_h$) was scaled with respect to the calculation time. 
When running GDRR for 5 minutes, for example, $\mathcal{L}_h$ was set to half of the value described in Section \ref{section:parameters}.

\section{Conclusion} \label{section:conclusion}
This paper introduced GDRR, a heuristic for solving the 2D bin packing problem with guillotine constraints.
Variants of the problem with a heterogeneous set of bins (variable-sized) and 90$\degree$ rotation of items are also supported.
The heuristic contains a ruin and recreate procedure which iteratively attempts to improve the solution. The search can be described as being goal-driven since it continuously strives to create feasible solutions with an ever decreasing limit of available bin area. 
To escape local optima, GDRR employs the Late Acceptance Hill-Climbing metaheuristic.
Unlike most other heuristics for this problem, the focus lies primarily on the improvement phase rather than the constructive aspect.
It is capable of consistently producing the best results in terms of solution quality across several benchmarks in the literature.

Future research may include exploring the 2BP variant without guillotine constraint and other cutting and packing problems with real-world constraints. 
Finally, it would also be interesting to investigate the applicability of a goal-driven approach in other types of problems.

\section*{Acknowledgment}
Editorial consultation provided by Luke Connolly (KU Leuven). The authors wish to thank the anonymous reviewers and the editor whose comments significantly improved the quality of the paper.

\bibliography{bibliography}

\begin{thebibliography}{}

\bibitem[Berkey and Wang, 1987]{berkey1987two}
Berkey, J.~O. and Wang, P.~Y. (1987).
\newblock Two-dimensional finite bin-packing algorithms.
\newblock {\em Journal of the operational research society}, 38(5):423--429.

\bibitem[Bortfeldt, 2006]{bortfeldt2006genetic}
Bortfeldt, A. (2006).
\newblock A genetic algorithm for the two-dimensional strip packing problem with rectangular pieces.
\newblock {\em European Journal of Operational Research}, 172(3):814--837.

\bibitem[Bresina, 1996]{Bresina1996}
Bresina, J.~L. (1996).
\newblock {Heuristic-Biased Stochastic Sampling}.
\newblock {\em AAAI/IAAI}, 1:271--278.

\bibitem[Burke and Bykov, 2017]{burke2017late}
Burke, E.~K. and Bykov, Y. (2017).
\newblock The late acceptance hill-climbing heuristic.
\newblock {\em European Journal of Operational Research}, 258(1):70--78.

\bibitem[Charalambous and Fleszar, 2011]{Charalambous2011}
Charalambous, C. and Fleszar, K. (2011).
\newblock {A constructive bin-oriented heuristic for the two-dimensional bin packing problem with guillotine cuts}.
\newblock {\em Computers and Operations Research}, 38(10):1443--1451.

\bibitem[Christiaens and Vanden~Berghe, 2020]{Christiaens2020}
Christiaens, J. and Vanden~Berghe, G. (2020).
\newblock Slack induction by string removals for vehicle routing problems.
\newblock {\em Transportation Science}, 54(2):417--433.

\bibitem[Clautiaux et~al., 2013]{clautiaux2013new}
Clautiaux, F., Jouglet, A., and Moukrim, A. (2013).
\newblock A new graph-theoretical model for the guillotine-cutting problem.
\newblock {\em INFORMS Journal on Computing}, 25(1):72--86.

\bibitem[Coffman et~al., 1980]{coffman1980performance}
Coffman, Jr, E.~G., Garey, M.~R., Johnson, D.~S., and Tarjan, R.~E. (1980).
\newblock Performance bounds for level-oriented two-dimensional packing algorithms.
\newblock {\em SIAM Journal on Computing}, 9(4):808--826.

\bibitem[Coffman~Jr and Shor, 1990]{coffman1990average}
Coffman~Jr, E. and Shor, P. (1990).
\newblock Average-case analysis of cutting and packing in two dimensions.
\newblock {\em European Journal of Operational Research}, 44(2):134--144.

\bibitem[Cui et~al., 2015]{Cui2015}
Cui, Y.-P., Cui, Y., and Tang, T. (2015).
\newblock {Sequential heuristic for the two-dimensional bin-packing problem}.
\newblock {\em European Journal of Operational Research}, 240(1):43--53.

\bibitem[Cui et~al., 2018]{Cui2018}
Cui, Y.-P., Yao, Y., and Zhang, D. (2018).
\newblock {Applying triple-block patterns in solving the two-dimensional bin packing problem}.
\newblock {\em Journal of the operational research society}, 69(3):402--415.

\bibitem[Dyckhoff, 1990]{Dyckhoff1990}
Dyckhoff, H. (1990).
\newblock A typology of cutting and packing problems.
\newblock {\em European Journal of Operational Research}, 44(2):145--159.

\bibitem[Fleszar, 2013]{fleszar2013three}
Fleszar, K. (2013).
\newblock Three insertion heuristics and a justification improvement heuristic for two-dimensional bin packing with guillotine cuts.
\newblock {\em Computers \& Operations Research}, 40(1):463--474.

\bibitem[Friesen and Langston, 1986]{friesen1986variable}
Friesen, D.~K. and Langston, M.~A. (1986).
\newblock Variable sized bin packing.
\newblock {\em SIAM journal on computing}, 15(1):222--230.

\bibitem[Gilmore and Gomory, 1965]{gilmore1965multistage}
Gilmore, P. and Gomory, R.~E. (1965).
\newblock Multistage cutting stock problems of two and more dimensions.
\newblock {\em Operations research}, 13(1):94--120.

\bibitem[Hong et~al., 2014]{Hong2014}
Hong, S., Zhang, D., Lau, H.~C., Zeng, X., and Si, Y.~W. (2014).
\newblock {A hybrid heuristic algorithm for the 2D variable-sized bin packing problem}.
\newblock {\em European Journal of Operational Research}, 238(1):95--103.

\bibitem[Hopper and Turton, 2002]{hopper2002problem}
Hopper, E. and Turton, B. (2002).
\newblock Problem generators for rectangular packing problems.
\newblock {\em Stud. Inform. Univ.}, 2(1):123--136.

\bibitem[Iori et~al., 2020]{iori2020exact}
Iori, M., de~Lima, V.~L., Martello, S., Miyazawa, F.~K., and Monaci, M. (2020).
\newblock Exact solution techniques for two-dimensional cutting and packing.
\newblock {\em arXiv preprint arXiv:2004.12619}.

\bibitem[Kröger, 1995]{kroger1995guillotineable}
Kröger, B. (1995).
\newblock Guillotineable bin packing: A genetic approach.
\newblock {\em European Journal of Operational Research}, 84(3):645--661.

\bibitem[Lodi et~al., 2002a]{lodi2002two}
Lodi, A., Martello, S., and Monaci, M. (2002a).
\newblock Two-dimensional packing problems: A survey.
\newblock {\em European journal of operational research}, 141(2):241--252.

\bibitem[Lodi et~al., 1999]{lodi1999heuristic}
Lodi, A., Martello, S., and Vigo, D. (1999).
\newblock Heuristic and metaheuristic approaches for a class of two-dimensional bin packing problems.
\newblock {\em INFORMS Journal on Computing}, 11(4):345--357.

\bibitem[Lodi et~al., 2002b]{Lodi2002}
Lodi, A., Martello, S., and Vigo, D. (2002b).
\newblock {Recent advances on two-dimensional bin packing problems}.
\newblock {\em Discrete Applied Mathematics}, 123(1-3):379--396.

\bibitem[Lodi et~al., 2017]{Lodi2017}
Lodi, A., Monaci, M., and Pietrobuoni, E. (2017).
\newblock {Partial enumeration algorithms for Two-Dimensional Bin Packing Problem with guillotine constraints}.
\newblock {\em Discrete Applied Mathematics}, 217:40--47.

\bibitem[Martello et~al., 2003]{martello2003exact}
Martello, S., Monaci, M., and Vigo, D. (2003).
\newblock An exact approach to the strip-packing problem.
\newblock {\em INFORMS journal on Computing}, 15(3):310--319.

\bibitem[Ntene and van Vuuren, 2009]{ntene2009survey}
Ntene, N. and van Vuuren, J.~H. (2009).
\newblock A survey and comparison of guillotine heuristics for the 2d oriented offline strip packing problem.
\newblock {\em Discrete Optimization}, 6(2):174--188.

\bibitem[Oliveira et~al., 2019]{ORdatasets}
Oliveira, O., Gamboa, D., and Silva, E. (2019).
\newblock Resources for two-dimensional (and three-dimensional) cutting and packing solution methods research.
\newblock {\em In Proceedings of the 16th International Conference on Applied Computing}, 53:131--138.
\newblock \url{https://github.com/Oscar-Oliveira/OR-Datasets}.

\bibitem[Ortmann et~al., 2010]{ortmann2010new}
Ortmann, F.~G., Ntene, N., and Van~Vuuren, J.~H. (2010).
\newblock New and improved level heuristics for the rectangular strip packing and variable-sized bin packing problems.
\newblock {\em European Journal of Operational Research}, 203(2):306--315.

\bibitem[Pisinger and Sigurd, 2005]{pisinger2005two}
Pisinger, D. and Sigurd, M. (2005).
\newblock The two-dimensional bin packing problem with variable bin sizes and costs.
\newblock {\em Discrete Optimization}, 2(2):154--167.

\bibitem[Polyakovsky and M’Hallah, 2009]{Polyakovsky2009}
Polyakovsky, S. and M’Hallah, R. (2009).
\newblock {An agent-based approach to the two-dimensional guillotine bin packing problem}.
\newblock {\em European Journal of Operational Research}, 192(3):767--781.

\bibitem[W{\"{a}}scher et~al., 2007]{Wascher2007}
W{\"{a}}scher, G., Hau{\ss}ner, H., and Schumann, H. (2007).
\newblock {An improved typology of cutting and packing problems}.
\newblock {\em European Journal of Operational Research}, 183(3):1109--1130.

\bibitem[Wei et~al., 2013]{wei2013goal}
Wei, L., Oon, W.-C., Zhu, W., and Lim, A. (2013).
\newblock A goal-driven approach to the 2d bin packing and variable-sized bin packing problems.
\newblock {\em European Journal of Operational Research}, 224(1):110--121.

\end{thebibliography}

\end{document}